\documentstyle[
	subeqnarray
	,aasms4
	,12pt
	]{article}

\begin{document}

\title{ 
Molecular Hydrogen Formation
on Astrophysically Relevant Surfaces
}
\author{
        { N.~Katz,$^1$ I.~Furman,$^1$ O.~Biham,$^1$ 
        V. Pirronello,$^2$ and G. Vidali$^3$\\
        $^1$Racah Institute of Physics, The Hebrew University of Jerusalem,
        Jerusalem 91904, Israel\\
        $^2$Istituto di Fisica, Universita' di Catania,
        95125 Catania, Sicily, Italy\\
        $^3$Department of Physics, Syracuse University,
        Syracuse, NY 13244
        }
}

\begin{abstract}

Recent experimental results about the formation of
molecular hydrogen on astrophysically relevant surfaces under 
conditions
close to those
encountered in the interstellar medium 
are analyzed using rate equations.
The parameters of the rate equation
model are fitted to temperature-programmed desorption 
curves obtained in the laboratory.
These parameters are the activation energy barriers for atomic
hydrogen diffusion and desorption, 
the barrier for molecular hydrogen
desorption, and the probability of spontaneous desorption of a hydrogen
molecule upon recombination.  
The model is a generalization of the Polanyi-Wigner equation
and provides a description of both first and
second order kinetic processes within a single model. 
Using the values of the
parameters that fit best the experimental results, 
the efficiency of
hydrogen recombination on olivine and amorphous carbon surfaces 
is obtained for a range of hydrogen flux and surface temperature 
pertinent to a wide range of interstellar conditions.
\end{abstract}
\keywords{dust--- ISM; abundances --- ISM; molecules --- molecular processes}


\baselineskip = 16pt

\section{Introduction}
        \label{intro.}

The formation of molecular hydrogen in the interstellar medium (ISM)
is a process of fundamental importance
(\cite{Duley1984,Williams1998}).  
It was recognized long ago (\cite{Gould1963}) 
that ${\rm H}_2$ cannot form in the gas phase 
efficiently enough to account for its abundance.
It was proposed that dust grains act as catalysts
allowing the protomolecule to quickly release 
the 4.5 eV of excess energy in a time comparable to the vibration
period of the highly vibrationally excited state in which it is formed.
The  problem can be described as follows.
An H atom approaching the surface of a grain has
a probability $\xi$ to become adsorbed.
The adsorbed H atom (adatom)
spends an average time $t_{\rm H}$ (residence time)
before leaving the surface.
If during the residence time the H adatom encounters another H adatom, 
an ${\rm H}_2$ molecule will form with a certain probability.

This problem has been studied theoretically over the years and
different models have been proposed
(\cite{Gould1963};
\cite{Williams1968};
\cite{Hollenbach1970};
\cite{Hollenbach1971a};
\cite{Hollenbach1971b};
\cite{Smoluchowski1981};
\cite{Aronowitz1985};
\cite{Duley1986};
\cite{Pirronello1988};
\cite{Sandford1993};
\cite{Takahashi1999};
\cite{Farebrother1999}).
In particular, Hollenbach et al.
calculated sticking and accommodation of 
H atoms in a semiclassical way, while the mobility was treated quantum 
mechanically. 
They concluded that tunneling between adsorption sites, 
even at 10K, would have assured the required mobility.
The steady state production rate of molecular hydrogen per unit volume
was expressed according to (\cite{Hollenbach1971b})

\begin{equation}
\label{eq:salpeter}
	R_{\rm H_2} = {1 \over 2}
		n_{\rm H} v_{\rm H} \sigma \gamma n_{\rm g},
\end{equation}

\noindent
where 
$n_{\rm H}$ 
and 
$v_{\rm H}$ 
are the number density and the speed
of H atoms in the gas phase, respectively,
$\sigma$ is the average cross-sectional area of a grain and
$n_{\rm g}$ 
is the number density of dust grains.
The parameter
$\gamma$ is the fraction of H atoms striking the grain
that eventually form a molecule, namely $\gamma = \xi \eta$,
where
$\eta$ 
is the probability that an H adatom on the surface
will recombine with another H atom
to form 
${\rm H}_2$.
The probability $\xi$ for an H atom to become adsorbed on a grain  surface 
covered by an ice mantle has been calculated by Buch and Zhang 
(1991) and Masuda et al. (1998).
Eq.\ (\ref{eq:salpeter})
states that, for $\eta = 1$, whenever two H atoms are adsorbed on a grain,
a ${\rm H}_2$ molecule is formed.

Recently, 
a series of experiments were conducted 
to measure
hydrogen recombination
in an ultra high-vacuum (UHV) chamber by
irradiating the sample 
with two beams of H and D atoms  
and monitoring the HD production rate
(\cite{Pirronello1997a,Pirronello1997b,Pirronello1999}).
The two beams were used in order to obtain a better signal to 
noise ratio than would have been possible for H$_2$.
Two different substrates have been used: 
a natural olivine (a silicate made of a mixture of 
Mg$_2$SiO$_4$ and Fe$_2$SiO$_4$) slab; 
and an amorphous carbon sample.
These samples are considered as better analogues of 
interstellar dust than any other model surface which was studied before. 
The substrate temperatures were in the range 
between 5K and 15K.
The HD formation rate was measured using 
a quadrupole mass spectrometer
both 
{\it during} 
and 
{\it after} 
irradiation with H and D atoms. 
In the latter case, a Temperature Programmed Desorption
(TPD) experiment was carried out in which the temperature of the
sample was quickly ramped to over 30 K
to desorb all weakly adsorbed species.

The main results obtained by Pirronello et al. (1997a, 1997b, 1999) 
are as follows:
(a)
In the temperature range of interest for interstellar applications
(between 10K and 15K),
the formation rates deduced from the experimental data are about 	   
one order of magnitude lower than those calculated by 
and Hollenbach \& Salpeter (1970, 1971) and Hollenbach et al. (1971);
(b)
According to the desorption spectra, hydrogen, that is adsorbed as atomic,
appears to acquire significant 
mobility only around 9K in the case of olivine 
and at a somewhat higher temperature
in the case of amorphous carbon,
even in the high coverage regime.
Thus, at temperatures lower than about 10K tunneling 
alone does not provide enough mobility to H adatoms to enable 
recombination, and thermal activation is required.

In this paper we perform a detailed analysis of the 
hydrogen recombination experiments 
of Pirronello et al.
(1997a, 1997b, 1999)
using a rate equation model.
The parameters of the rate equation
model are fitted to the experimental TPD curves.
These parameters are
the activation energy barriers for atomic
 hydrogen diffusion and desorption, 
the barrier for molecular hydrogen
desorption and the probability of spontaneous desorption of a hydrogen
molecule upon recombination.  
Using the values of the
parameters that fit best the experimental results, 
the efficiency of
hydrogen recombination on the olivine and amorphous carbon surfaces 
is obtained for a range of 
hydrogen fluxes 
and surface temperatures  
pertinent
to a wide range of interstellar conditions.

The paper is organized as follows. 
In Sec. 
\ref{sec:Experimental} 
we describe the experiments to be analyzed. The rate equation model is 
introduced in Sec. 
\ref{sec:Model}. 
Subsequent analysis and results are
presented in Sec. 
\ref{sec:Analysis}, 
followed by a discussion
in Sec. 
\ref{sec:Discussion} 
and a summary in Sec. 
\ref{sec:Summary}.
 
\section{Review of Experimental Methods}

\label{sec:Experimental}

The experimental apparatus and measurement techniques 
are described in 
Pirronello et al. 
(1997a, 1997b, 1999)
and in greater detail in 
Vidali et al. 
(1998a). 
Here we give a brief outline. 
The apparatus  consists of 
an ultra-high vacuum (UHV) chamber pumped by a cryopump and a
turbomolecular pump 
(operating pressure in the low 10$^{-10}$ torr range).  
The sample is placed in the center of the UHV chamber and
mounted on a liquid helium continuous flow cryostat. 
By varying the
flow of liquid helium and with the use of a heater located behind the
sample, temperatures can be maintained in the range of 5-30 K. For
cleaning purposes, the temperature of the sample can be raised to
about 200$^\circ$C (without liquid helium in the cryostat). 
The temperature
is measured by an iron-gold/chromel thermocouple and a calibrated
silicon diode placed in contact with the sample.  Two triple
differentially pumped atomic beam lines are aimed at the surface of
the sample.  Each has a radio-frequency cavity in which the molecular
species is dissociated, cooled to $\sim$ 200 K by passing the atoms
through a cooled Al channel, and then injected into the
line. Dissociation rates are typically in the 75 to 90$\%$ range, and
are constant throughout a run. 
Estimated fluxes are as low as 10$^{12}$ atoms 
cm$^{-2}$s$^{-1}$ 
(\cite{Vidali1998a}).

The reason for using two different
lines and two isotopes (one line for H and the other for D) is 
that in preliminary runs using only one line, it became
evident that the signal of H$_2$ formation was hidden in the
background given by the undissociated fraction of molecules coming
directly from the beam source. 
The possibility of using a second line
is undoubtedly one of the most important features of this equipment.
By using H atoms in one line and D atoms in the other, we can look at
the formation of HD on the surface, knowing that there are no other
spurious sources of HD. The signal of HD is collected by a quadrupole
mass spectrometer mounted on a rotatable flange. The experiment is
done in two phases. First, H and D beams are sent onto the surface for a
given period of time (from tens of seconds to tens of minutes).
At this time any HD formed and released is detected. 
In the second phase
(the TPD phase), the
sample temperature is quickly ($\sim$ 0.6 K/sec) ramped and the HD
signal is measured. 

By measuring the desorption rate 
$R(t)$ 
as a function of time, 
as well as the temperature of the surface as a
function of time, information on the kinetics of desorption can be
obtained. The desorption rate 
can by approximated by the Polanyi-Wigner equation:

\begin{equation}
R(t) = \nu N(t)^{\beta} \exp(- E_d / k_B T), 
\label{eq:desorption}
\end{equation}

\noindent
where $N$ is the number density of
reactants on the surface, 
$\beta$ is the order of desorption, 
$\nu$ is the attempt frequency,
$E_d$ is the effective
activation energy for the dominant recombination and desorption process
and $T=T(t)$ is the sample temperature.
In the TPD experiment,
first order 
($\beta=1$)
desorption curves 
$R(t)$
exhibit asymmetric peaks with a sharp 
drop-off on the right hand side. 
The position of the peak is insensitive to coverage.
Second order desorption curves ($\beta=2$) 
exhibit symmetric peak shapes.
These peaks shift towards
{\em lower temperatures} 
as coverage is increased
(\cite{Chan1978}). 

In the experiments analyzed here the irradiation stage 
was done at a surface
temperature between $T=5K$ and $T=7K$,
for several irradiation time intervals. 
The HD desorption rate vs. surface temperature
during the TPD runs is shown in Figs. 1 and 2 for olivine
and in Figs.
3 and 4 for amorphous carbon.
The TPD curves shown in Figs. 1 and 3 exhibit first order
kinetics due to the larger irradiation times, 
while the ones shown in Figs. 2 and 4
exhibit second order kinetics.
For both the olivine and the amorphous carbon samples, 
at the sample temperature examined here, 
most of the HD detected is formed because of
thermal activation during the heat pulse. 
Only a small fraction of HD
is formed during the irradiation process, showing that, at least under
our experimental conditions, prompt-reaction mechanisms 
(\cite{Duley1986}) 
or fast tunneling 
(\cite{Hollenbach1971b}) 
are not that important.

\section{Model}
\label{sec:Model}

\subsection{Assumptions}

In the desorption curves studied here most of the adsorbed hydrogen is 
released well before a temperature of $30$K is reached. 
Therefore, we assume that
the hydrogen atoms on the surface are trapped in 
physisorption potential wells and are thus only weakly adsorbed. 
We also assume that the mechanism for the creation of
H$_{2}$ (or HD) is 
the Langmuir-Hinshelwood (LH) scheme, 
namely that the rate of creation
of H$_{2}$ is diffusion limited. 
This assumption is justified
due to the observed Langmuir-like 
kinetics of the amounts desorbed as a 
function of the irradiation time 
(\cite{Pirronello1997a}).
Since the coverages involved in the experiments analyzed here
are low, other mechanisms such as the Eley-Rideal (ER) scheme, 
in which hydrogen atoms, coming from the gas phase, collide and promptly react
with already adsorbed hydrogen atoms, 
are of less importance and are not included in our model. 
Furthermore, the experimental results indicate that the ER mechanism does
not contribute significantly, since even at higher
coverages there was little desorption of HD during the irradiation phase. 

The model we present here, which reproduces quite well the experimental
desorption curves, and the choice of the assumptions on which it is based
are the results of other less successful attempts.
We have, in fact, tried to fit the desorption curves using a model in which
the H and D populations of adatoms that are used in the experiments
were described separately. 
In these earlier attempts we assumed that all HD molecules are
promptly released upon formation, as investigations 
on more regular metal surfaces
suggest 
(\cite{Rettner1996};
\cite{Winkler1998}). 
The model in that case has four parameters:
the diffusion barriers as well as the
desorption barriers for H and D adatoms.
We found that such model cannot provide a reasonable fit to the
experimental desorption curves.

In the model we present here, 
we do not assume spontaneous desorption once
a molecule is created on the surface 
despite the $\sim4.5$ eV released in the recombination. 
In order not to increase the
number of parameters to be used in the fits, we decided not to treat
separately the two populations of H and D adatoms, but to consider only one
population of H adatoms characterized by average properties. In this way we
kept the number of parameters to four: 
the activation energy
barriers for diffusion and desorption of hydrogen adatoms, 
the activation energy
barrier for desorption of molecular hydrogen that had not been released
into the gas phase upon formation, and the probability $1-\mu$ 
of spontaneous desorption of a hydrogen molecule upon
recombination.

An important final assumption is that
all energy barriers are coverage independent. 
This assumption may not apply at high coverage. 
However, at the low coverages obtained in the experiments
analyzed here (up to $\sim 1\% $ of a layer),
it is a reasonable assumption.

\subsection{Rate Equations}
\label{sec:Rate Equations}

Consider an experiment in which a flux of H atoms 
is irradiated on the surface. 
H atoms that stick to the
surface, 
once the surface temperature is raised,
perform hops as random walkers with increased frequency
and recombine when they encounter one another.
Let $N_{1}(t)$ [in monolayers ($ML$)] be the coverage 
of H atoms on the surface and $N_{2}(t)$ (also in $ML$) 
the coverage of H$_{2}$ molecules. 
We obtain the following set of rate equations:

\begin{subeqnarray}
\label{eq:N}
\dot{N_{1}} & = & F(1 - N_{1} -  N_{2}) - P_{1}N_{1} - 2\alpha N_{1}^{2} 
\slabel{eq:N1} \\
\dot{N_{2}} & = & \mu \alpha N_{1}^{2} - P_{2} N_{2}. 
\slabel{eq:N2}
\end{subeqnarray}

\noindent
The first term on the right hand side of 
Eq.~(\ref{eq:N1}) 
represents the incoming
flux in the Langmuir kinetics. 
In this scheme H atoms deposited on top of H atoms
or H$_{2}$ molecules already on the surface are rejected. 
$F$ represents an
{\em effective} flux (in units of $ML s^{-1}$), 
namely it already includes the possibility of a temperature
dependent sticking coefficient. 
The second term in 
Eq.~(\ref{eq:N1}) 
represents the desorption of H atoms from the
surface. 
The desorption coefficient is 

\begin{equation}
P_{1} =  \nu \cdot \exp (- E_{1} / k_{B} T)  
\label{eq:P1}
\end{equation}

\noindent
where $\nu$ is the attempt rate 
(standardly taken to be $10^{12}$ $s^{-1}$), 
$E_{1}$ 
is the activation energy barrier for desorption 
of an H atom and $T$ is the temperature.
The third term in 
Eq.~(\ref{eq:N1}) 
accounts for the depletion of the H population
on the surface due to recombination into H$_{2}$ molecules, 
where

\begin{equation} 
\alpha =  \nu \cdot \exp (- E_{0} / k_{B} T) 
\label{eq:Alpha}
\end{equation}

\noindent
is the hopping rate of H atoms on the surface
and $E_{0}$ is the activation energy barrier for H diffusion. 
Here we assume that there is no barrier for recombination. If such a barrier 
is considered, it can be introduced as discussed in Pirronello et al. (1997b, 
1999).
The first term on the right hand side of 
Eq.~(\ref{eq:N2}) 
represents the creation of H$_{2}$ molecules. 
The factor $2$ in the third term of 
Eq.~(\ref{eq:N1}) 
does not appear here since it
takes two H atoms to form one molecule. 
The parameter
$\mu$ represents the fraction of H$_{2}$ molecules
that remains on the surface upon formation, 
while a fraction of $(1-\mu)$ is spontaneously desorbed due 
to the excess energy released in the recombination process.
The second term in 
Eq.~(\ref{eq:N2}) 
describes the desorption of H$_{2}$ molecules. 
The desorption coefficient is 

\begin{equation}
P_{2}  =  \nu \cdot \exp (- E_{2} / k_{B} T), 
\label{eq:P2}
\end{equation}

\noindent
where $E_{2}$ is the activation energy
barrier for H$_{2}$ desorption.
The H$_{2}$ production rate $R$ is given by:  

\begin{equation}
R  =  (1-\mu) \cdot \alpha N_{1}^{2} + P_{2} N_{2}. 
\label{eq:Production}
\end{equation}

\noindent
This model can be considered as a generalization of 
the Polanyi-Wigner model 
[see 
Eq. 
(\ref{eq:desorption})].
It gives rise to a wider 
range of simultaneous applications, 
compared to 
Eq. 
(\ref{eq:desorption}).
In particular, it
describes both first order 
and second order 
desorption kinetics 
(or a combination) 
for different regimes of temperature and flux. 

In the experiments analyzed here, 
both the temperature and the flux were controlled
and monitored throughout. 
Each experiment consists of two phases. 
In the first phase
the sample temperature is constant up to time $t_0$, 
under a constant irradiation rate $F_0$. 
In the second phase, the irradiation is turned off and
linear heating of the sample is applied at the rate 
$b$ ($K s^{-1}$):
 
\begin{eqnarray}
\label{eq:temp}
F(t) & = & F_{0}; \ \ \ \ \ \  T(t)  =  T_{i}: 
\ \ \ \ \ \ \ \ \ \ \ \ \ \ \ \ \ \ \ 0 \leq t < t_{0}  
\slabel{eq:temp1} \\
F(t) & = & 0;    \ \ \ \ \ \ \ T(t)  =  T_{i} + b (t - t_{0}): 
\ \ \ \ \  t \geq t_{0}.
\slabel{eq:temp2}
\end{eqnarray} 

\noindent
Here  $T_{i}$ is the constant temperature of 
the sample during irradiation. 

In the case that the rejection terms in  
$F(1-N_{1}-N_{2})$ 
are neglected and 
the effective
flux becomes simply $F$ 
(a valid assumption at low coverages), 
the rate equations can be solved analytically.  
However, the solution is expressed in terms of
intractable nested integral expressions, 
and is of little use to us. 
In the case we study here, in which the 
rejection terms are taken into
account, no such solution exists and the 
equations are integrated numerically.

\section{Analysis and Results}
\label{sec:Analysis}

\subsection{Methods}
\label{subsec:Methods}

We will now examine to what extent the rate equation
model can describe the experimental results. 
To this end we performed 
numerical integration of 
Eqs.~(\ref{eq:N1})-(\ref{eq:N2}) 
with the aid of a Bulirsch-Stoer stepper algorithm 
(\cite{Press1992}).
The result of the integration is a set of 
TPD curves that are a function of the
chosen set of parameters.
A standard TPD experimental run includes 
the time dependence of the flux 
$F(t)$
and temperature
$T(t)$ as well as the four parameters
$E_{0}$, $E_{1}$, $E_{2}$ and $\mu$.
The temperature $T(t)$ is measured directly via a thermocouple. 
The flux $F(t)$ 
($s^{-1}$) 
is estimated as described elsewhere 
(\cite{Vidali1998b}). 
An approximate value for 
$F(t)$, 
in the required units of $ML s^{-1}$, 
can be obtained by integrating the TPD spectra, 
generating the total {\it yield} 
of the various experiments. The flux is then obtained from 
the exponential fit indicated by Langmuir kinetics. 
It is important to stress that
this is a lower bound value for the flux, and this value is reached 
only if there is no H desorption at all. 
We are now left with the four parameters 
$E_{0}$, $E_{1}$, $E_{2}$ and $\mu$
which are assumed to be independent 
of the flux or temperature.
These parameters form a four dimensional space that has to be explored 
in order to find the values for
which the calculated TPD curves provide 
the best fit to the experimental TPD ones.
  
The merit function to be minimized in the fitting procedure is 
the standard $\chi^{2}$ function, which is the
sum over the squares of the differences between the experimental points and
the calculated ones.
Another possibility that we considered was to compare the derivatives 
of the experimental TPD curves with the ones of the
simulation (again using $\chi^{2}$). 
This possibility required using the Savitzki-Golay filtering
(\cite{Press1992}) 
in order to obtain a reasonably smooth 
derivative from the experimental data, 
because of the well known increase in the noise to signal 
ratio in any derivative of measured data.

To obtain the parameters that best fit the experimental data
one needs to probe the parameter space and find 
the set of values of the parameters that give rise to
the global minimum of the merit function. 
The probing procedure we used is based on random search.
A point in the four dimensional parameter space is
randomly picked. 
A numerical integration of the model's equations is 
performed with the chosen parameters.
The merit function obtained from the comparison of the 
resulting curve with the experimental
curve
is then evaluated. 
If this set represents an improvement over the current minimum, 
the new point is accepted
as the new minimum. 
If, however, the new set did not score lower (i.e. better) than the known 
minimum, it was promptly discarded. The probability distribution 
for picking the next
random set of parameters was taken to be a Lorentzian centered 
around the current set of
parameters and thus favored nearby points.  
The relatively slow drop-off of the Lorentzian
function also allowed the occasional taking of longer steps
and thus prevented the process from getting stuck
in a local minimum.

\subsection{Results}
\label{subsec:Results}

The experimental TPD curves and the fits obtained by the 
rate equations are shown in
Figs. 1 and 2 for the olivine sample and in 
Figs. 3 and 4 for the amorphous carbon sample. 
The parameters obtained in the fitting procedure 
for the two samples are shown in Table 1. 
The parameter set generated for each sample 
represents a simultaneous best fit for 
all the six TPD curves.
Fitting each curve separately typically 
produce better fits, but at the expense of an increased range in the 
values of the parameters. 
These variations allow us to generate approximate error estimates. 
It is found that
the energy barriers 
$E_{0}$ and $E_{2}$ are
very well determined by this process 
(to within several tenths of a meV). 
The barrier
$E_{1}$ is not as
well determined, and its values given above, for both samples, 
are to be taken as lower bounds 
to the correct value (within 3 meV). 
The parameter
$\mu$ is determined to within $ \pm 0.1$, and thus
justifies our assumption, within this model, that not all
H$_{2}$ molecules immediately 
desorb upon recombination.
Attempting to artificially force $\mu = 0$ 
and do the fits  with the remaining three
parameters, degrades the fit substantially 
and cannot recreate the entire
range of behavior of the data simultaneously.

Although 
Eq.~(\ref{eq:desorption}) 
can be used to fit the entire range of experimental TPD results,
this equation, which includes a single activation energy, does not
provide as much insight as our model
[Eqs. (\ref{eq:N})].
For example, applying 
Eq.~(\ref{eq:desorption}) 
in the case
of Fig. 1 
(olivine, 1st order desorption kinetics), 
we use $\beta = 1$ and arrive at $E_{d} = 26.8 meV$. 
This is equivalent to the $E_{2}$ found for olivine using this model. 
However, as we shift to Fig. 2 (olivine, 2nd order desorption kinetics), 
we must now take $\beta = 2$, which results in $E_{d} = 24$ meV. 
Here the rate limiting process is the diffusion on the surface, 
and therefore $E_{d}$ is closer to $E_{0}$.
Unlike 
Eq.~(\ref{eq:desorption}),
our model
does not require setting a parameter (such as $\beta$).
Furthermore,
it provides the best-fit values of each of the relevant activation
energies for both the first and second order kinetics and using the same 
framework.

The model 
[Eqs. (\ref{eq:N})]
can also describe the steady state conditions 
which are reached when both the 
flux and the temperature are fixed. 
The steady state solution is then easily 
obtained by setting 
$\dot{N_{1}}$ 
and 
$\dot{N_{2}}$ to $0$ 
and solving the 
quadratic equation for $N_{1}$.
The coefficients 
$P_{1}$, $P_{2}$ and $\alpha$
in the rate equations are temperature
dependent and under the steady 
state assumption maintain constant values. 
The complete solution under these assumptions 
takes the form
(after neglecting the unphysical possibility of the negative root):  

\begin{subeqnarray}
\label{eq:steady}
N_{1} & = & \frac
{\sqrt{(P_{1}+F)^{2} + 8(\alpha F + \frac{\mu \alpha F^{2}}
{2P_{2}})} - (P_{1} + F)} {4(\alpha F + 
\frac{\mu \alpha F^{2}} {2P_{2}})}
\slabel{eq:steady1}	\\
N_{2} & = & \frac
{(P_{1}+F)^{2} + 4(\alpha F + \frac{\mu \alpha F^{2}}{2P_{2}}) - 
(P_{1}+F) \sqrt{(P_{1}+F)^{2} + 8(\alpha F + 
\frac{\mu \alpha F^{2}} {2P_{2}})}}
{8(\alpha + \frac{\mu \alpha F}{2P_{2}})^{2}}. 
\slabel{eq:steady2}
\end{subeqnarray}

\noindent
Note that when the rejection terms in the flux are neglected, 
the expressions for the steady 
state coverages and recombination rate 
become significantly simpler (\cite{Biham1998}).
These solutions may be useful in the study of 
recombination processes in the
interstellar medium
where steady state conditions may be relevant. 

In
this work, we present a model that
captures the kinetics of the 
diffusion-recombination-desorption process. 
The parameters thus obtained can then be used to study 
the astrophysically relevant cases. 
For example,  
by assuming steady state conditions, we can  
obtain the recombination efficiency as a function 
of flux $F$ and temperature $T$ for a range of parameters 
that goes from the astrophysically relevant 
(extremely low flux, $10-15$K) to the ones 
used in the laboratory (low flux, $5-30$K).
The recombination efficiency is defined as the ratio between
the production rate 
$R$ 
[Eq.~(\ref{eq:Production})] 
and the 
deposition rate $F/2$ (in molecules).

Varying $T$ and $F$ over the astrophysically relevant range we can 
identify the
regions in which 
there is non-negligible recombination 
efficiency. 
The recombination efficiency as a function of $T$ and $F$ 
is shown in Fig. 5 for olivine 
and in Fig. 6 for amorphous carbon.  
The main conclusion from these figures is that the  
recombination efficiency is highly temperature dependent. 
There is an efficiency window along the temperature axis
which shifts to
higher temperatures as the flux is increased. 
It is found that
under astrophysically relevant irradiation rates the
efficiency for olivine drops off at approximately $7 \sim 8$K,  
while for amorphous carbon it drops off only at 
$13 \sim 14$K. 

We find that, for both samples, the parameter $\mu$ 
(the probability of an H$_{2}$ molecule 
to remain on the surface upon recombination) 
has little effect on the 
production rate $R$ of molecular hydrogen
under steady state conditions.
This is easy to understand, since under 
steady state conditions the production rate
$R$ must be equal to the recombination rate on the surface,
and, thus, must be independent of $\mu$. 
The coverage of hydrogen molecules on the surface is
adjusted accordingly.
Similarly, the
energy barrier for desorption of molecular hydrogen, $E_{2}$, 
has little effect on the recombination efficiency,
under steady state conditions,
as long as it remains significantly smaller 
than the barrier for atomic desorption, $E_{1}$.

\section{Discussion}
\label{sec:Discussion}
  
Amorphous materials in general, and their surfaces in particular,
are difficult to characterize, due to their irregular
structure and composition.
In the experiments analyzed here, it gives rise to some uncertainty
about the role of quantum effects.
In principle, tunneling of H atoms should be considered.
However, the experiments indicate that
quantum effects appear to be small,
as the mobility
of the hydrogen atoms is very low at the lower irradiation temperatures 
(\cite{Pirronello1997b}).
Another issue is that in the experiments 
the desorption rates of HD were measured and the extrapolation
to processes involving H and H$_{2}$ is non-trivial.  
This difficulty is unavoidable, however, 
due to the large background noise in the measurements of H$_{2}$ production.
Careful calibration of the apparatus was used in generating the 
necessary extrapolation coefficients.
Taking isotopic effects into account in the rate equations is 
not easy for this type of surfaces, since the shape of the 
energy surface is not known. An attempt was made to introduce 
separate fitting parameters for H and D, while keeping $\mu=0$ 
(to maintain the same number of parameters). 
This fit was considerably worse than the ones presented here. 
Since isotopic effects enter as a shift in the activation 
energies of the various processes described here, it is fair 
to consider an ``effective'' hydrogen atom in the simulations.

Despite these obstacles, we 
obtained good fits to the
experimental TPD curves
using numerical integration of the rate equation model.  
This seems to indicate that most of the 
processes that occur at the microscopic level 
are captured by the dynamics of the
rate equations, or at least they average out to this global dynamics. 

According to common physical intuition based on long range 
forces and the values of atomic and molecular polarizabilities, 
H$_{2}$ is expected to be bound to the 
substrate more tightly than H.  
This, however, does not appear to be the case: 
the parameters calculated numerically, for both samples, put 
the H atoms in deeper wells than H$_{2}$.  
This feature must signal the fact that more complicated interactions 
(based also on shorter range forces) are taking place.

The recombination efficiency diagrams
(Figs. 5-7)
include two regions of low efficiency on both sides of the 
high efficiency window. 
The asymptotic 
steady state coverage in the region on the left hand side approaches 
unity, while everywhere else is very small. 
This is due to the fact that at such low temperatures
atoms remain stuck in place and barely diffuse; thus,  
recombination is inhibited.
These atoms accumulate on the surface blocking adsorption sites (``Langmuir 
rejection'').
However, this high coverage regime is not one that we 
can extrapolate to with any certainty within this rate equation model.
As already suggested in Pirronello et al. (1999),
mechanisms such as Eley-Rideal or diffusion by tunneling, which are 
not taken into account in the model,
may become significant in this regime.
Nevertheless, we can speculate that if the Langmuir rejection remains 
significant even at
higher coverages, the trend in recombination efficiency
shown should remain qualitatively correct.  
Luckily, such low temperatures are rarely of astrophysical 
interest.  
For olivine, at temperatures above $6.5$K, the 
calculated asymptotic coverage is very low and well within the regimes of 
experimentation and subsequent numerical
simulation; consequently, the relevance of the model is justified.

We find that the recombination efficiency on olivine
is high in the temperature range
of roughly $5-10$K; however, this temperature range
is lower than the one encountered in the interstellar clouds.
Therefore, we believe that this material is not a very 
likely candidate for an efficient 
catalyst of hydrogen recombination in interstellar space.

The recombination efficiency on amorphous carbon, on the other hand, 
behaves differently (recall Fig. 6). 
Here we see that the final drop in efficiency of $H_2$ recombination
is at higher temperatures, and amorphous carbon seems to be a more 
appropriate candidate of interstellar grains on which 
hydrogen may recombine with
high efficiency.

In a previous paper 
(\cite{Biham1998}) 
two limiting expressions were obtained for the 
H$_2$ production rate per unit volume under steady state conditions. 
Such expressions are linear (when $ \alpha F \gg P_{1}^{2}$)
or quadratic
(when $\alpha F \ll P_{1}^{2}$) in the flux of gas phase atoms.
Alternatively, they are independent
(when $ \alpha F \gg P_{1}^{2}$)
or quadratically dependent
(when $\alpha F \ll P_{1}^{2}$)
on the coverage of H adatoms.
Note that in Biham et al. (1998) the value $\mu=1$ was taken.
In that work, we studied the steady state behavior of the 
rate equations for the processes described here, 
where the value of $\mu$ 
does not affect the recombination 
efficiency.
The first of the two limits of
the steady state production rate of molecular hydrogen per unit volume
coincides with  the expression of Hollenbach's et al. 
[see Eq. (\ref{eq:salpeter})].
However, 
it is found that the second of the two limits, which is valid when there is a 
very low coverage of H adatoms on
interstellar grains, coincides with the expression proposed by Pirronello et al.
(1997b):

\begin{equation}
\label{eq:pirronello2}
        R_{\rm H_2} = 
                (n_{\rm H} v_{\rm H} \sigma \xi t_{\rm H})^2 n_{\rm g} 
                \alpha\gamma^{\prime},
\end{equation}
where $t_{\rm H} = 1/P_1$ and $\gamma^{\prime}$  
is the probability that two H adatoms recombine upon encountering
(taking into account 
that there might be an activation energy for recombination).

In Fig. 7 we present a contour diagram of the recombination 
efficiency on the carbon surface. 
On this diagram,
the parameter values for which 
$\alpha F = P_{1}^{2}$
are plotted as a
starred line. 
In the region on the left hand side
of the starred line, the calculated $R_{H_2}$
approaches the value  obtained when $ \alpha F \gg P_{1}^{2}$.
On the right hand side of the starred line, the calculated $R_{H_2}$
approaches the value obtained in the limit of $\alpha F \ll P_{1}^{2}$ .
The transition from the regime where $ \alpha F \gg P_{1}^{2}$ 
holds to the other, $\alpha F \ll P_{1}^{2}$, 
is quite rapid because of the 
exponential nature of the temperature dependence. 
We 
conclude that the two cases discussed in 
(\cite{Biham1998}) 
apply in a wide region of the diagrams (Fig.5-7).

The diagrams described above apply under conditions close to 
steady state. 
This is not necessarily
the case in the interstellar clouds. 
If the entire cloud is far from 
steady state, 
one must return to the original  
rate equation model and take into account the time dependence of the
flux and temperature.

The values obtained for $\mu$ 
imply a non-negligible probability for the hydrogen molecules 
to remain on the surface
instead of being immediately ejected into the gas phase. 
This result was found to be unavoidable within the assumption of the
model used here.
There are various mechanisms for efficient heat transfer from the molecule
to the surface. These may 
dissipate the excess energy and prevent immediate desorption. 
One possible mechanism 
may be due to the very irregular structure of the sample surfaces:
olivine was mechanically polished 
(hence there should be grooves of hundreds of 
Angstroms in width and depth). 
Atomic force microscopy (AFM) 
images of the olivine surface show such a rugged landscape 
at submicron scale, 
while amorphous carbon 
is composed of grains of the size of a hundred Angstrom. 
A definite possibility 
(more relevant in the amorphous carbon case, 
as the somewhat larger $\mu$ value suggests) 
is that hydrogen molecules, 
even if promptly released upon formation, 
do not necessarily go directly into the vacuum
but undergo a multiple series of collisions in which part of their
energy is released to the solid with a subsequent re-adsorption.
Such mechanism was identified for H and D atoms impinging on an
amorphous ice particle
(\cite{Buch1991}).
More experiments are needed in order to elucidate 
the nature of recombination
process at the atomic scale and to obtain directly a value for $\mu$.

\section{Summary}
\label{sec:Summary}

Experimental results on the formation of molecular 
hydrogen on various materials and 
conditions relevant to interstellar clouds were analyzed using rate
equations.
By fitting the results of a TPD experiment to a rate
equation model, four essential parameters of the process were obtained. 
These are the activation energy barriers for 
atomic hydrogen diffusion and  
desorption, the barrier for molecular hydrogen desorption, 
and the probability of spontaneous desorption of a hydrogen molecule upon
recombination. 
The results compare favorably with what is obtained 
from the Polanyi-Wigner equation.
Furthermore, the model represents a 
generalization that allows us to describe  
both first and second order processes 
(or even a combination of the two) within a single model.

In this work we have shown a procedure to extrapolate data taken 
under conditions available in the laboratory to values that should 
hold in astrophysical environments,
and to determine the efficiency of various 
surfaces as catalysts in molecule production. 
Polycrystalline olivine was shown to be 
inefficient as a catalyst in the relevant 
temperature/flux regime. On amorphous carbon 
a higher efficiency was reached due to higher
desorption/diffusion barriers that cause a 
rise in the recombination efficiency 
at the relevant temperatures. 

The proposed model and methods of analysis 
are neither complete nor definitive, 
but nonetheless represent a serious 
improvement over what has been used so far to predict the behavior
of hydrogen on astrophysically relevant surfaces.

\section{Acknowledgments}
\label{sec:Acknowledgments}

G.V. acknowledges support from NASA grant NAG5-6822.

\vspace{1in}
Table 1: The four parameters obtained by the 
fitting of the TPD curves for 
poly-crystalline olivine
and amorphous carbon. $E_{0}$ is the barrier for 
atomic diffusion, $E_{1}$ and $E_{2}$ are
the barriers for atomic and molecular desorption respectively, 
$1-\mu$ is the probability
of spontaneous desorption of a newly formed H$_{2}$ molecule.


\begin{tabular}{||c|c|c|c|c||} \hline
{\rm material} & {$\em E_{0}$(meV)} & {$\em E_{1}$(meV)} & 
{$\em E_{2}$(meV)} & {$\mu$} \\ \hline \hline
olivine & 24.7 & 32.1 & 27.1 & 0.33 \\ \hline
amorphous carbon & 44.0 & 56.7 & 46.7 & 0.413 \\ \hline
\end{tabular}

\newpage
\begin{center}
Figure Captions
\end{center}

\figcaption{TPD curves for higher coverage 
		experiments on an olivine slab. Irradiation 
	times are (in minutes) 8.0 ($ \protect\bullet $), 5.5 
($ \protect\times $) and 2.0 ($ + $). Fits
	are in solid lines.} 

\figcaption{TPD curves for lower coverage 
		experiments on an olivine slab. Irradiation 
	times are (in minutes) 0.55 ($ \protect\bullet $), 0.2 
($ \protect\times $) and 0.07 ($ + $). 
	Fits are in solid lines.} 

\figcaption{TPD curves for higher coverage 
		experiments on amorphous carbon. Irradiation 
	times are (in minutes) 32.0 ($ \protect\bullet $), 16.0 
($ \protect\times $) and 8.0 ($ + $). 
	Fits are in solid lines.} 

\figcaption{TPD curves for lower coverage 
		experiments on amorphous carbon. Irradiation 
	times are (in minutes) 4.0 ($ \protect\bullet $), 2.0 
($ \protect\times $) and 1.0 ($ + $). 
	Fits are in solid lines.} 

\figcaption{Recombination efficiency at steady state of the 
olivine slab as a function of $Log_{10}(F)$  
		(flux in $ML/sec$) and 
		T (temperature in $K$).}

\figcaption{Recombination efficiency at steady state of 
amorphous carbon 
as a function of $Log_{10}(F)$ 
		(flux in $ML/sec$) and 
		T (temperature in $K$).}

\figcaption{Contour plot of the recombination 
efficiency at the steady state for amorphous carbon, as a 
	function of $Log_{10}(F)$ (flux in $ML/sec$) and T 
(temperature in $K$). 
	The starred line represents $\alpha F = P_{1}^{2}$. }
    
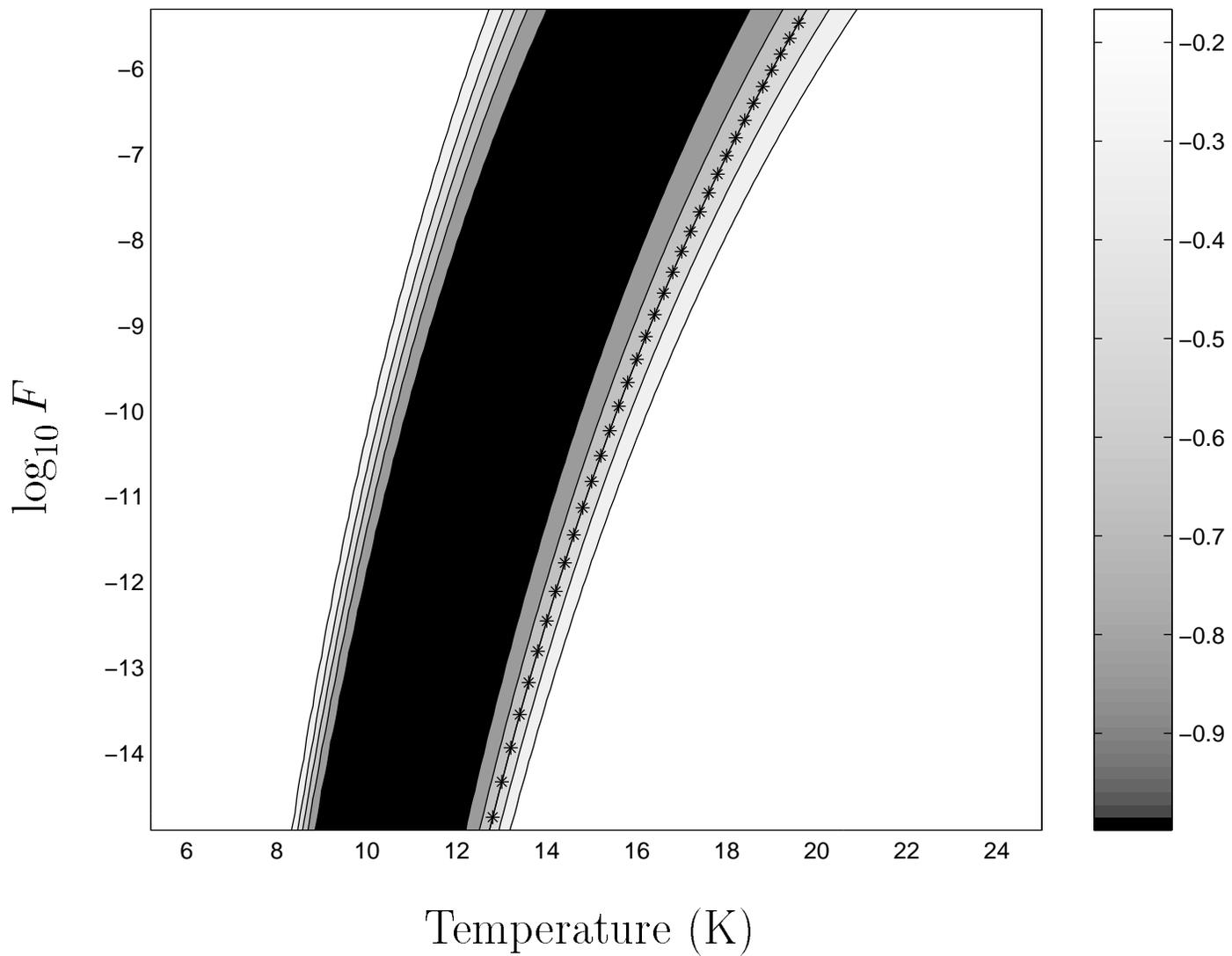
\end{document}